\documentclass{ws-ijmpc}

\begin{document}
\font\note=cmr8

\markboth{Authors' Names} {}

\catchline{}{}{}{}{}

\title{TOWARDS A STATISTICAL PHYSICS OF HUMAN MOBILITY}

\author{GALLOTTI RICCARDO, BAZZANI ARMANDO, RAMBALDI SANDRO}

\address{Department of Physics, University of Bologna and INFN Bologna\\ via Irnerio
46, 40126 Bologna, Italy, bazzani@bo.infn.it}

\maketitle

\begin{history}
\received{Day Month Year}
\revised{Day Month Year}
\end{history}

\begin{abstract}
In this paper, we extend some ideas of statistical physics to
describe the properties of human mobility. From a physical point of
view, we consider the statistical empirical laws of private cars
mobility, taking advantage of a GPS database which contains a
sampling of the individual trajectories of $2\%$ of the whole
vehicle population in an Italian region. Our aim is to discover
possible "universal laws" that can be related to the dynamical
cognitive features of individuals. Analyzing the empirical trip
length distribution we study if the travel time can be used as
universal cost function in a mesoscopic model of mobility. We
discuss the implications of the elapsed times distribution between
successive trips that shows an underlying Benford's law, and we
study the rank distribution of the average visitation frequency to
understand how people organize their daily agenda. We also propose
simple stochastic models to suggest possible explanations of the
empirical observations and we compare our results with analogous
results on statistical properties of human mobility presented in the
literature. 
\end{abstract}

\ccode{PACS Nos.: 11.25.Hf, 123.1K}

\section{Introduction}

Human mobility  has recently become a fruitful research field in
complexity science\cite{vespignani2012}, as it offers the
possibility of performing the study of a complex system at many
description levels. Indeed, Information Communication Technologies
allow to collect big databases on the individual dynamics and make
human mobility a paradigmatic example of a statistical system of
cognitive particles\cite{bazzani2009}. Understanding human mobility
in urban contexts has relevant consequences not only for urban
planning\cite{rozenfeld2008}, opinion spreading\cite{vespignani2012}
and epidemic dynamics\cite{colizza2006,hufnagel2004}, but it
contributes to the formulation of a statistical mechanics for
complex systems. In our opinion this last goal requires three
fundamental steps:
\begin{itemize}
\item to discover universal statistical laws that describe the
average properties of human mobility;
\item to characterize the microscopic dynamics of individuals sharing the same urban environment;
\item to study the existence of critical phenomena (i.e. phase
transitions) due to the individual interactions in presence of
limited resources or the appearance of transient states driven by
external sudden changes.
\end{itemize}
Up to now, the scientific efforts have been focused on making
progress in the first two items, whereas the third item seems to be
beyond the possibilities of the present scientific methodologies.
The existence of statistical laws is based on the definition of
macroscopic observables that give information on the global state of
the system (e.g. the definition of a mobility temperature). The
study of microscopic individual dynamics would aim to understand the
mobility strategies, i.e. to discover common features in the use of
time and space and in the organization of mobility agenda  that are
related to cognitive behaviors.  In the paper\cite{brockmann2006}
the individual mobility is recorded indirectly by tracking the
movements of dollar bills and a power-law distribution for spatial
displacements have been proposed. In a second paper, the individual
mobility has been analyzed by using access data to specific internet
sites\cite{noulas2011}.  In these cases we have an indirect measure
of human mobility without details on individual behavior, due to
non-biunivocal correspondence between performed activities and the
use of bills or the internet accesses. Other papers study the
individual mobility using the localization of the mobile phone
calls\cite{gonzalez2008,bagrow2012} and GPS (Global Position System)
taxi database\cite{bin2009,liang2012}. To work with mobility data
the relation between the type of activity and the frequency of the
phone calls must be understood or the cost of the taxi service must
be considered. In all cases the study of human mobility within urban
contexts is difficult due to the nature of urban mobility, dominated
by short trips. In this paper we use a GPS database from vehicles
collected in Italy. Approximately, $2\%$ of the whole vehicle
population is monitored by insurance reasons and time, position,
velocity and covered distance are recorded by sampling each
trajectory at a spatial scale of 2 km or a time scale of 30 seconds.
Moreover, a datum is also recorded each time the engine is switched
on or off. Even if one has no control on the population sample, the
database offers a unique possibility to study the human mobility at
a fine spatial scale on large urban areas using long time series. We
analyze the GPS data recorded in the whole Emilia Romagna region
during the month of November 2009. We have filtered the data to
consider only trips whose initial and end points are inside the
region and only individuals that mainly live and perform their
mobility in the considered area.  The spatial extension of Emilia
Romagna is approximately $220\times 100$ km$^2$ and it allows to
study both small scale urban mobility and intercity mobility. Our
aim is to develop a statistical physics approach to human mobility
based on universal properties and/or cost
functions\cite{domencich1975,kolb2003}, that characterize the
individual behavior at mesoscopic level. This is performed by
studying the main statistical features that describe the dynamical
properties of mobility and the relation with the downtimes: i.e. the
time intervals between two successive trips, during which the GPS
system is switched off. We assume that a location can be associated
to each downtime. The paper is organized as follows: in the first
section we present the GPS database used by our analysis, in the
second section we discuss the statistical laws related to the use of
space and time. In the third section we study the statistical
properties that could be consequences of individual cognitive
behaviors.

\section{GPS Data: preprocessing and main properties}
\begin{figure}[h]
\centerline{\psfig{file=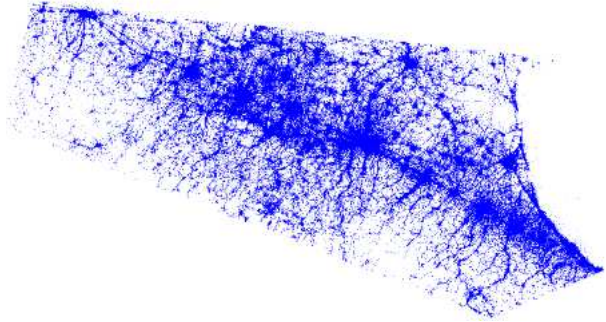,width=8 truecm}} \vspace*{8pt}
\caption{Georeferentiatied GPS data recorded in the Emilia Romagna
region during the month November 2009. The data distribution allows
to detect the main cities locations along the ancient \textit{Via
Emilia} that crosses longitudinally the region, the geometry of the
main road network and the population spread in the territory.
\label{dataER}}
\end{figure}

The GPS database is collected by a private company (Octo Telematics
s.p.a.) for insurance reasons and refers to $\simeq 2\%$ of the
whole vehicle population. Due to the italian law on privacy we have
no direct information on individuals, but the installation of a GPS
system on a vehicle entitles the holder to a discount off the
insurance price. This is particular appealing for young people so
that it expected a bias in this sense in the considered sample. The
taxi companies or the delivering services use their own GPS systems
and they do not contribute to the database, which is mainly set up
by private vehicles. There is a small percentage of vehicles used
for professional reasons and belonging to private companies, that
take advantage of the insurance discounts of collective contracts.
In our analysis we have selected people whose mobility is performed
inside the Emilia Romagna region, discarding the trips whose origin
or destination are outside the region, so that we expect to describe
the private car mobility performed by citizens living inside the
considered region. The data set refers to a sample of about $75,000$
monitored vehicles that complete 7.7 million trips internal to the
region. A GPS measure is recorded each time the engine is switched
on or off and the trips are sampled each 2 km or 30 seconds
depending on the central system needs. The recorded data are time,
position (longitude and latitude), actual velocity, covered distance
from the initial measure and GPS quality signal for each
vehicle\cite{bazzani2011}. A filtering procedure has been applied to
discard trajectories affected by systematic errors that prevent the
correct vehicle georeferencing on the road network, or the
evaluation of trip length or duration\cite{bazzani2011}. The
expected error in a typical GPS measure is about $10\; m$ in the
position, whereas it is negligible in the time. In the figure
\ref{dataER}, we plot the Georeferenced GPS data recorded during the
whole month of November 2009 inside the Emilia Romagna region.
The data distribution allows to recognize the location of the main
cities along the historical axis from east to west defined by the
roman road \textit{Via Emilia}. The most part of the population
lives in the vicinity of the ancient \textit{Via Emilia} or in the
northern part of the region, where the \textit{Padana} plain
stretches towards the \textit{Po} river. The southern part of the
region is mainly mountainous and barely populated. On the eastern
part we have the Adriatic sea coast, that attracts tourists with
entertainment activities. From a physical point of view the vehicles
realize a dynamical system on a configuration space defined by a
network structure and the microscopic dynamics should reflect the
individual strategies. Indeed the complexity of urban environments
do not allow to reduce the mobility problem to the origin
destination paradigm ruled by circadian rhythms\cite{portugali2012}.
But each individual seems to organize independently his mobility,
trying to minimize the interaction with other individuals. As a
consequence the role of "free will" becomes relevant and we have
significant stochastic effects in a mesoscopic description. In the
next section we try to characterize the statistical properties of
individual mobility.

\section{A statistical physics approach}

As starting point we consider how to justify some of the typical
assumptions of classical statistical physics\cite{landau1980}. Human
mobility can be seen a dynamical system constrained by the road
network and driven by the individual mobility demand. The road
network is organized in a hierarchical way, according to the spatial
scale and the relevancy of traffic flow, from the highways to the
small urban streets. Moreover, the sprawling
phenomenon\cite{portugali2012}, that influences the development of
modern cities, implies an increasing contribution of non-systematic
mobility. To synthesize our interpretation of experimental
observations, we formulate some a priori assumptions on the
individual behavior.\par Firstly, the individuals behave as almost
independent particles, since each person organizes the mobility
according to his propensities. This means that individuals interact
due to traffic rules and that the relevance of collective mobility
phenomena is expected to be small in average (of course there can be
exceptional events).
\par
Secondly, habits have strong influence in human mobility.  Our
experimental data suggest that the individuals tend to repeat the
same path when going from the same origin to the same destination.
Moreover, many locations are visited several times during the month.
This also means that, in normal conditions, the dynamical system
associated to human mobility reaches a stationary state (mainly true
for working days; weekends are different).\par Thirdly, at the base
of the individual mobility there is a set of strategies and,
according to transport engineers\cite{wardrop1952}, the travel-time
seems to play the role of cost function.
\par
\begin{figure}[h]
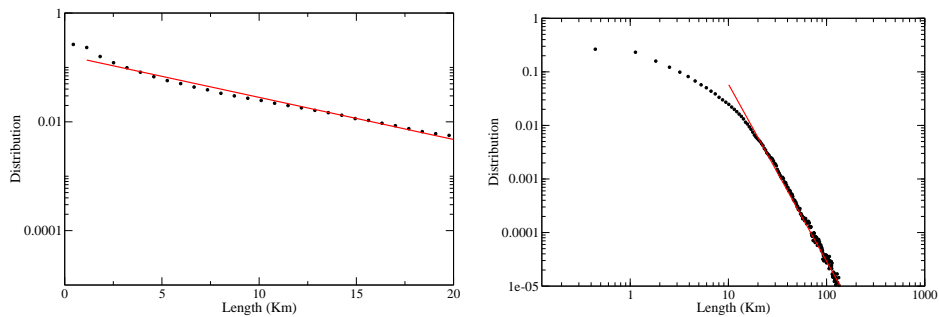

\centerline{\psfig{file=trip_len_lin.eps,width=6 truecm}\quad
\psfig{file=trip_len_log.eps,width=6 truecm}} \vspace*{8pt}
\caption{Statistical distribution of the trip lengths for the whole
Emilia Romagna computed by using GPS data recorded during the
November 2009: we use log-lin scale in the left picture and log-log
scale in the right picture. The log-lin scale a possible exponential
behavior for the short trips that corresponds to $95\%$ of the data
(red line). The log-log scale points out a possible interpolation of
the distribution tail by a power law: $p(l)\propto l^{-3.3}$ (red
straight line). \label{trip}}
\end{figure}
\begin{figure}[h]
\centerline{\psfig{file=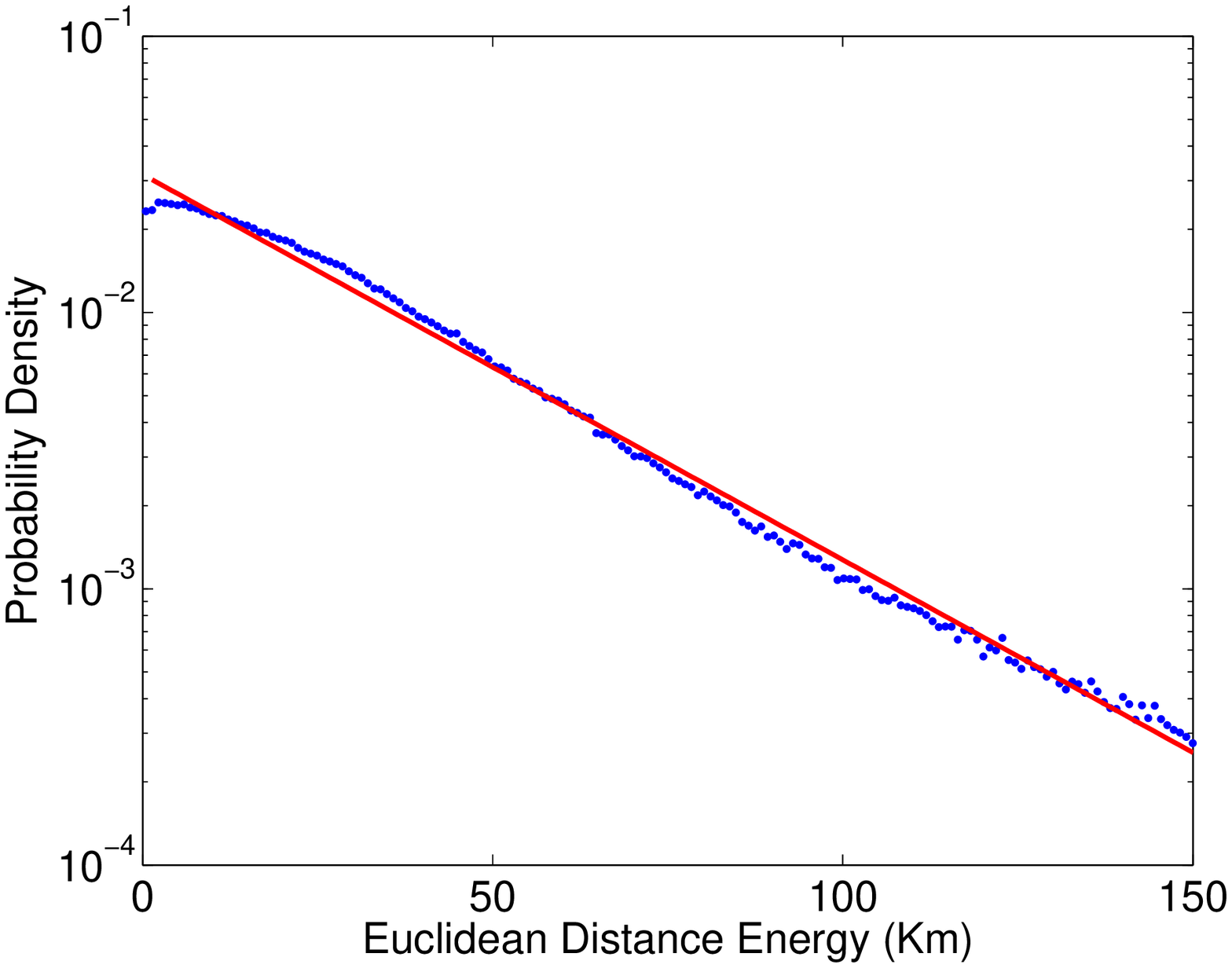,width=6 truecm}
\quad\psfig{file=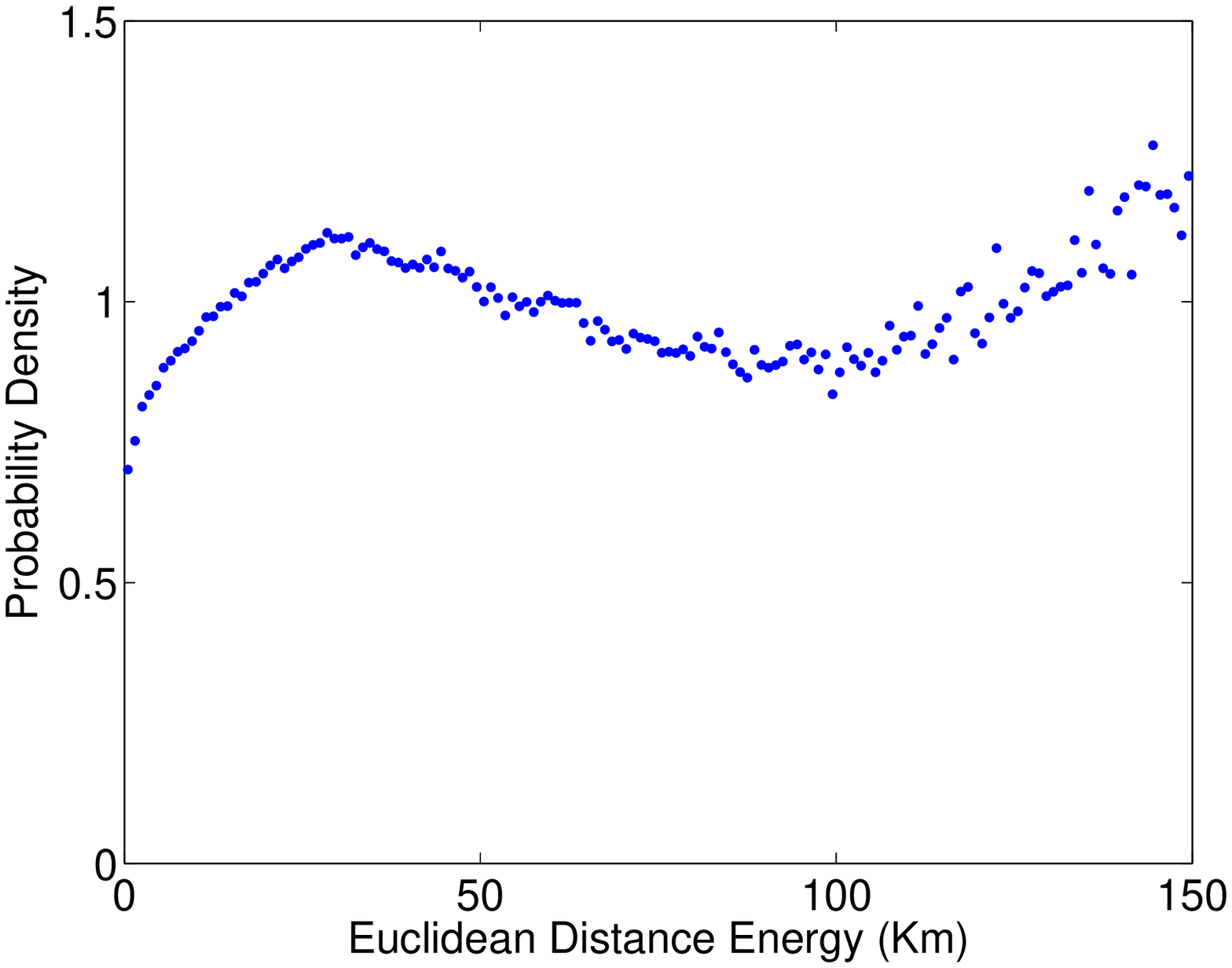,width=6 truecm}}
\vspace*{8pt} \caption{(Left plot) Daily mobility distribution from
the GPS data selecting people moving inside the Emilia Romagna
region. The straight line refers to an exponential fit of the
distribution with a characteristic length $\beta^{-1}=30.4\pm 0.4$
m. (Right plot) $m(\lambda)$ distribution (cfr. definition
(\ref{mdef})) computed using (\ref{MB1}). \label{moben}}
\end{figure}

We also remark that the previous assumptions should not be
considered a description of a individual microscopic dynamics, but a
theoretical framework to develop a statistical physics approach by
means of mesoscopic models. Under this point of view, individuals
can be represented as particles moving between some preferred
destinations (home, workplace, ...), which are chosen with a daily
periodicity, with a stochastic dynamics due to traffic rule and
vehicle interactions. To support this picture we look for empirical
statistical laws, inferred from the GPS data, consistent with the
previous hypotheses. We start considering the trip lengths
distribution (figure \ref{trip}), computed using the GPS data. The
trip length is defined by the covered distance on the road network
computed integrating the path length from the location where the
engine is switched on up to the location where the engine is
switched off (the trip length is recorded by the GPS system). We
consider a trip completed when the rest time is longer than 5
minutes otherwise we sum the lengths between two successive stops.
We remark on three main features:
    \begin{romanlist}[(b)]
    \item the very short trips ($l\le 2$ km) have a great statistical relevance;
    \item there exists a characteristic trip length $\simeq 6.2$ km;
    \item the long trip distribution recalls a fat tail (power law) distribution.
    \end{romanlist}
The trip length distribution reflects the way everybody realizes his
mobility demand in connection with the spatial activity
distribution\cite{simini2012}. We propose a theoretical explanation
for the distribution using the previous hypotheses. As a consequence
of the circadian rhythms, it is quite natural to consider the daily
mobility $\lambda$ of each individual defined by the sum of the trip
lengths of performed within an interval of 24 hours. We expect the
existence of an average daily mobility for the individuals both for
physical and economical reasons (any trip has a cost in time and
energy). The daily mobility distribution computed from the GPS data
is plotted in fig. \ref{moben} together with an exponential
interpolation.
We suggest a theoretical explanation for this behavior by dividing
the territory into a number of different locations $x\in X$, with
homogeneous geographical features. Assuming a given activity
distribution in the territory, we associate to each location a daily
mobility length $\lambda_x$ defined by the average distance that an
individual has to cover each day to satisfy his mobility demand (in
other words $\lambda_x$ measures the accessibility of the
$x$-location to the existing activities). Let $p_x$ be a priori
probability that an individual chooses to live in the $x$-location
without taking into account any mobility cost\footnote{In a
homogeneous territory $p_x$ would be constant, otherwise $p_x$ may
depend on the geographical features.}. Assuming that individuals act
as independent particles, the probability associated to a
distribution $\{n_x\}$, where $n_x$ is the number of individuals in
the location $x$, is given by a multinomial distribution
\begin{equation}
w(\{n_x\})=\prod_x \left (\frac{p_x^{n_x}}{n_x!}\right
)\label{weight}
\end{equation}
Applying a maximal entropy principle with the constraints that the
total number of individuals and the total mobility are finite
$$
\sum_x n_x=N\qquad \sum_x \lambda_x n_x=\Lambda
$$
one can determine the most probable distribution. Maximizing the
Gibbs entropy\cite{landau1980}
\begin{equation}
S=-\sum_x w(n_x)\ln w(n_x)  \label{entropy}
\end{equation}
we get the Maxwell-Boltzmann distribution
\begin{equation}
\rho(x)=A\exp(-\beta \lambda_x)p_x \label{MB}
\end{equation}
where $A$ is a normalizing constant and $\beta$ depends on the
average mobility $\beta^{-1}=\Lambda/N$. Adding over all the
locations with the same value $\lambda_x=\lambda$, we finally get
the distribution
\begin{equation}
\rho(\lambda)=A\,m(\lambda)\exp(-\beta \lambda) \label{MB1}
\end{equation}
where
\begin{equation}
m(\lambda)=\sum_{\lambda_x=\lambda}p_x \label{mdef}
\end{equation}
The measure $m(\lambda)$ gives the statistical weight of individuals
that would perform a daily mobility $\lambda$, if their distribution
in the territory would not depend on mobility costs. As shown in
fig. \ref{moben}, the daily mobility distribution is quite well
interpolated by an exponential distribution in the interval $10\;
km<\lambda< 150$ km. The distribution $m(\lambda)$ estimated
according to the formula (\ref{MB1}) (figure \ref{moben} right), has
a limited variation within this interval with a local maximum at
$\lambda\simeq 30$ km, that reflects the macroscopic spatial
distribution of activities in the Emilia Romagna territory.
Therefore a possible explanation for the $m(\lambda)$ behavior is
the following: considering that the activities are mainly located in
the cities, the initial increase of $m(\lambda)$ is due to the
population living in the attraction basin of the cities and the
maximum at $\simeq 30$ km gives an estimate of the average distance
among the main cities.\par
\begin{figure}[h]
\centerline{\psfig{file=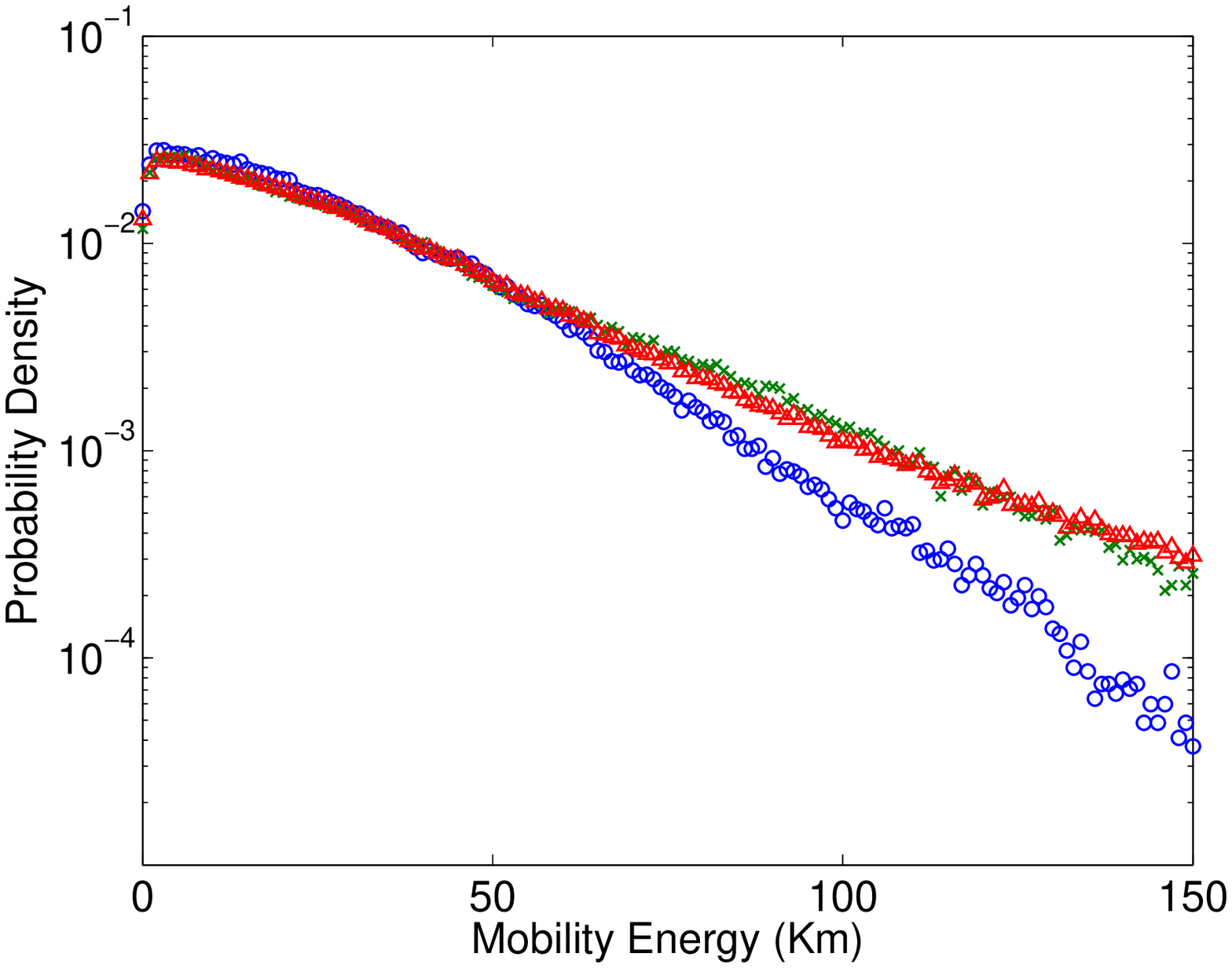,width=6 truecm}\quad
\psfig{file=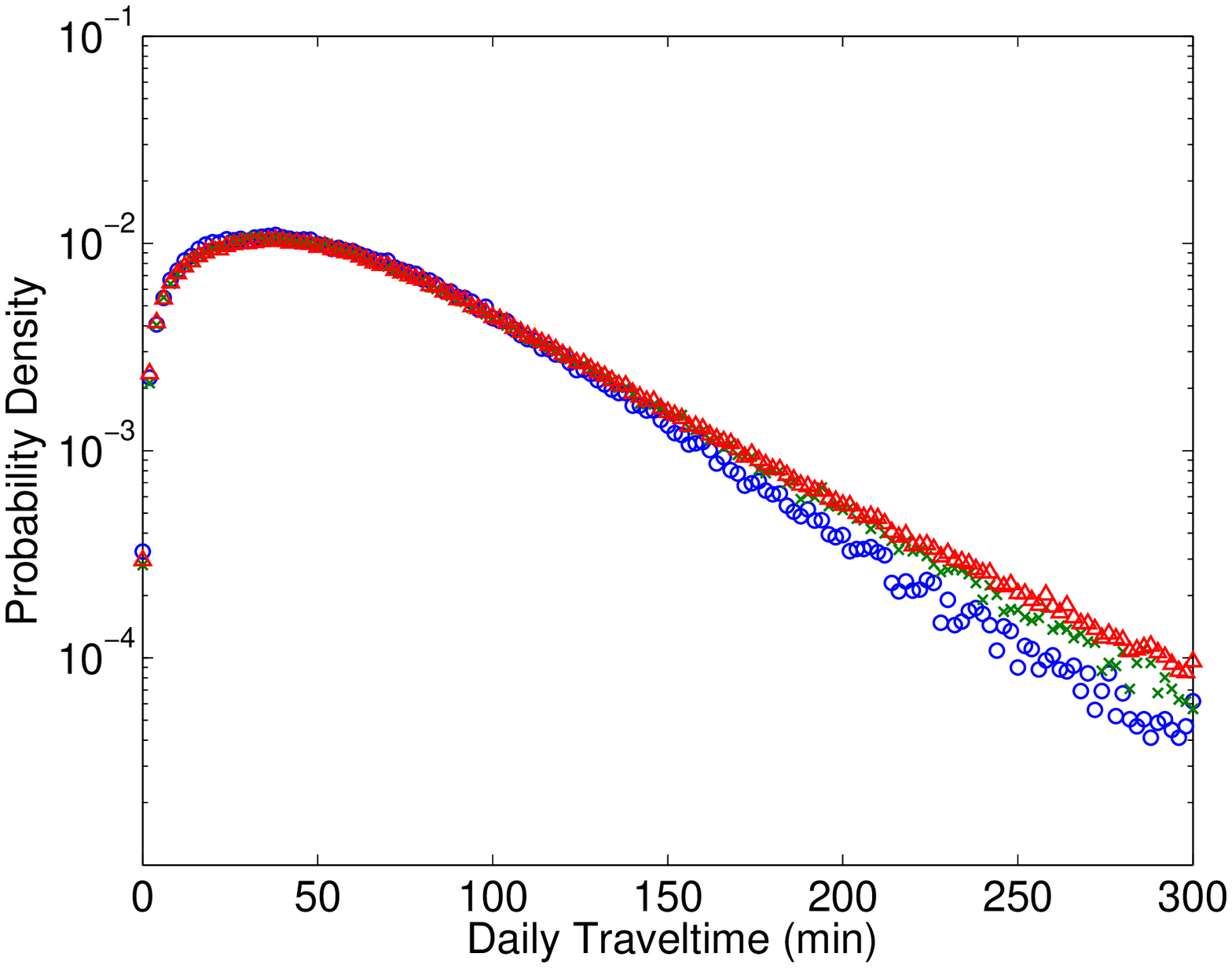,width=6 truecm}}
\vspace*{8pt} \caption{(Left picture) Daily mobility distributions
computed considering individuals which perform their mobility inside
regions of different size around the Bologna center: the circles
refer to the Bologna province $R\simeq 30$ km, the crosses refer to
region that includes the nearby cities $R\simeq 50$ km and the
triangles give the distribution for the whole region. (Right
picture) Daily travel time distribution corresponding to the daily
mobility distributions plotted in the left picture: the different
symbols have the same legenda as in the left picture and refer to
the same areas.\label{moben_area}}
\end{figure}

 The statistical distribution (\ref{MB})
leaves open the question if the exponential decay is related to the
extension of the considered region. We, then, have compared the
daily mobility related to areas of different size $R$ centered
around Bologna (the regional capital), from the Bologna province
($R\le 30$ km), to the area enclosing the nearby cities ($R\le 50$
m) and then to the whole region. In each area we have only
considered individuals whose mobility is performed internally to the
area itself, but that have not been previously (for smaller radius)
considered. We recall that our analysis refers to the use of private
vehicles and we expect that cars are used to satisfy the same
mobility demand in all the cases; this is false inside urban areas
($R<5$ km) where one has a good availability of public means and
more restriction in the use of private cars. The resulting
distributions are reported in fig. \ref{moben_area} where the
exponential decaying can be clearly detected at different scales,
and for large daily mobility we see a different behavior close to
the main city.
The results suggest that an entropic principle is robust in
describing the average mobility demand, but the characteristic
spatial scale decreases approaching an urban area. Considered the
travel time distributions (see figure \ref{moben_area} (right)),
interestingly, they tend to collapse into a single curve. This is an
experimental evidence that time may define an universal cost for
mobility (once the transportation mean is given): the average
mobility time is estimated $70$ minutes from the GPS
data\footnote{This value could be interpreted as the daily time that
an individual accepts to invest in his mobility.}.\par
\begin{figure}[h]
\centerline{\psfig{file=speedvariance.eps,width=6 truecm}\quad
\psfig{file=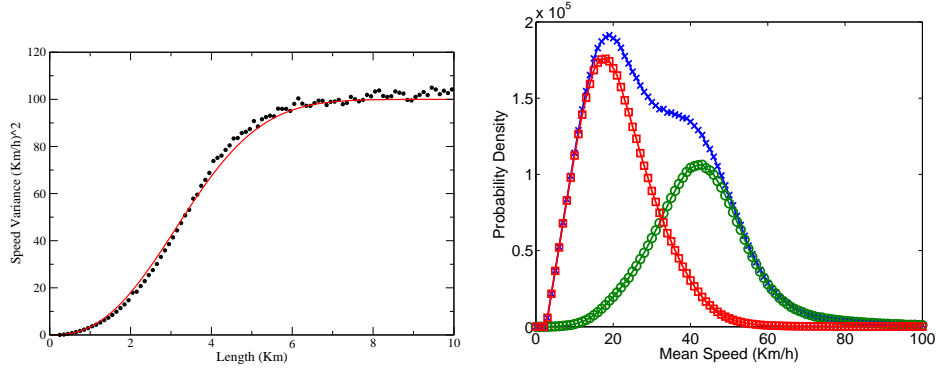,width=6 truecm}} \vspace*{8pt}
\caption{(Left picture) Average speed variance as a function of the
trip length: we have computed the average speed for a given trip
length using the GPS data of the whole Emilia-Romagna with a
discretization step of 100 $m$. The continuous line is a data
interpolation using the function
$\sigma^2(l)=\sigma_0^2(1-\exp{-(l/b)^{5/2}})$ with $\sigma_0=10$ km
and $b=3.8$ km. (Right picture) Average speed distribution for the
recorded trips (blue curve): the distribution can be decomposed into
the sum of two distributions considering the trips whose length is
$\le 5$ km (red curve) and the remaining ones (green curve).
\label{speed}}
\end{figure}

\begin{figure}[h]
\centerline{\psfig{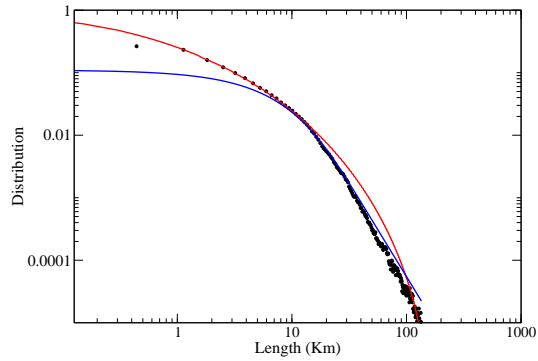}}
\vspace*{8pt} \caption{Statistical distribution of the trip lengths
measured using GPS data (black dots): the red curve refers to the
distribution (\ref{distril1}), whereas the blue curve is computed
using eq. (\ref{distril2}) with $\lambda=180$ km and $n=30$.
\label{trip1}}
\end{figure}

 Comparing the
figures \ref{moben_area} left and right, we remark that the
space-time relation cannot be reduced to a simple proportionality.
The reason is twofold: from one hand there is an intrinsic
heterogeneity in the human mobility due to different drivers
behaviors, on the other hand the small scale structure of the road
network influences the vehicle dynamics. To better understand the
space-time relation, we have studied the variance of the average
speed as a function of the trip length. In the figure \ref{speed} we
plot the result for the whole GPS data set: the data show a power
law increase of the variance for very short trips and a relaxation
to a stationary condition for trip lengths greater than 8 km: the
stationary variance corresponds to a \textit{rms} $\sigma_0\simeq
10$ km/h in the speed distribution (the red line show an
interpolation of the experimental data).
However, considering the average speed distribution for the whole
trips, it is possible to point out two different typologies: the
short trips $l\le 5$ km with an average velocity $\simeq 20$ km/h
and a small variance and the longer trips with a distribution
centered at $\simeq 45$ km/h and with a rms $\sigma_0\simeq 10$ km/h
(fig. \ref{speed} left). The velocity distribution for the short
trips probably refers to urban mobility, but the correlation of the
profile with the road network features is an open problem. We remark
that the two trip typologies are not directly related to the
exponential and power law behavior in the trip length distribution
(see fig. \ref{trip1})since the power law behavior can be detected
considering trip longer than 20 km.
\par
\begin{figure}[h]
\centerline{\psfig{file=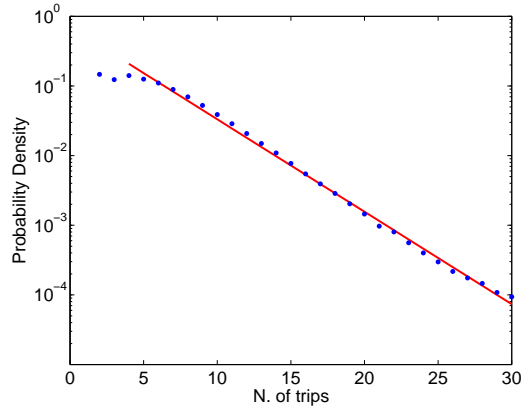,width=7.
truecm}}\vspace*{8pt} \caption{Distribution of the daily activity
number for the sampled individuals in Emilia Romagna; the continuous
line is an exponential interpolation $p(n)\propto \exp(-n/a)$ with
$a=3.27\pm .08$. \label{activity}}
\end{figure}

\noindent
In order to relate the trip length distribution (fig. \ref{trip})
with the daily mobility (fig. \ref{moben_area}), we consider how
many trips each individual makes in a day.  In figure \ref{activity}
we plot the probability distribution of the trip number together
with an exponential interpolation.
For $n\le 5$ we have about half of the sample population that
performs a systematic mobility, whereas when $n>5$ the exponential
decay suggests a statistical equilibrium without any particular
structure in the individual mobility. To interpret the statistical
part of the trip distribution (cfr. fig.\ref{trip}), we consider an
ensemble of particles characterized by a total mobility $\lambda$,
and for each particle we randomly distribute at most $n$
destinations within the interval $[0,\lambda]$. The obtained
distances among the destinations are the trips performed by
individual-particles. Given $\lambda$ and $n$, the trip distribution
can be computed analytically according to\cite{bazzani2010}
\begin{equation}
p_{n,\lambda}(l)\propto \sum_{k=1}^n
e^{-k/a}(k+1)k(1-l/\lambda)^{k-1} \label{distril}
\end{equation}
where $l\in [0,L]$ and $a=3.27$ is determined from the trip
distribution (see fig \ref{activity}). Using the exponential
distribution (\ref{MB1}, where we neglect the changes due to
$m(\lambda)$, and integrating over $\lambda$, we get an analytic
formula for the trip distribution
\begin{equation}
p_n(l)\propto \int_{\lambda_m}^{\lambda_M} p_{n,\lambda}\exp(-\beta
\lambda) d\lambda \label{distril1}
\end{equation}
In the fig. \ref{trip1} we compare the empirical trip distribution
with our analytical result (\ref{distril1}).
Formula (\ref{distril1}) closely interpolates the experimental data
for short trip lengths $l\le 15$ km \footnote{The discrepancy at
very small trip $l<1$ km is expected since using the exponential
distribution $e^{-k/a}$ for the activities, we have overestimated
the small trips.}. This allows to reproduce the mobility of $\simeq
87\%$ of the observed users, that correspond to $\simeq 52\%$ of the
total space traveled. But we have a discrepancy in the tail of the
empirical distribution. A possible explanation is obtained if one
does not introduce the exponential decaying in the number of trips
(cfr. fig. \ref{activity}), so that the distribution (\ref{distril})
reads
\begin{equation}
p_{n,\lambda}(l)={2\lambda^2\over n(n+3)}\sum_{k=1}^n
(k+1)k(1-l/\lambda)^{k-1}= {2\lambda^3\over n(n+3)}{d^2\over
dl^2}(1-l/\lambda)^2{\displaystyle 1-(1-l/\lambda)^n\over l}
\label{distril2})
\end{equation}
Since $1-l/\lambda$ is small for long trip ($l\simeq \lambda$) for
$n\gg 1$ we can approximate
\begin{equation}
p_{n,\lambda}(l)\simeq {2\lambda^3\over n(n+3)}{d^2\over
dl^2}(1-l/\lambda)^2{1\over l}\propto l^{-3}+O(\lambda^{-2})
\label{distril3}
\end{equation}
The power law tail (\ref{distril3}) seems to be in agreement with
the empirical observations (see fig. \ref{trip1} for a comparison of
(\ref{distril2}) with the experimental data). This result suggests
that we have users with a number of trips higher than the
statistical expectation and with a large mobility: probably they use
the vehicle for working reasons. We remark that the power law
$p(l)\propto l^{-3}$ is different from the power-laws suggested by
the dollar bill displacement distribution $p(l)\propto
l^{-1.59}$\cite{brockmann2006} or by the mobile phone data
$p(l)\propto l^{-1.75}$\cite{song2010}. But both consider a much
larger spatial scale and do not refer to a particular transportation
mean. Conversely it is consistent with the taxi data
analysis\cite{liang2012}, that is related to human mobility in a
large road network. The analysis suggests that the complexity of
human mobility is due to strong interactions among individuals or to
environment structure, but it is mainly due to the individual
mobility organization in space and time.

\section{Time and human activities}

\begin{figure}[h]
\centerline{\psfig{file=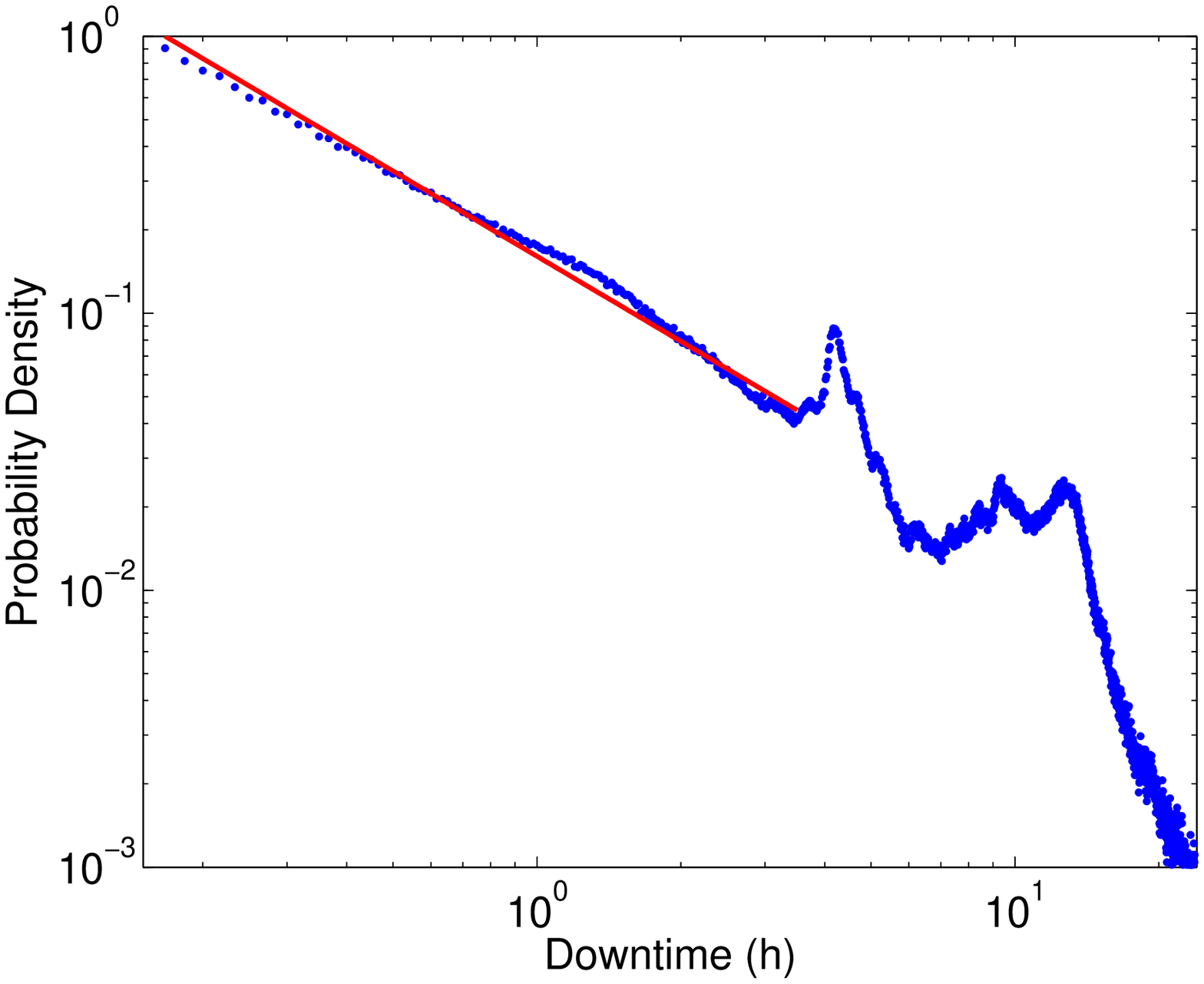,width=6 truecm}\quad
\psfig{file=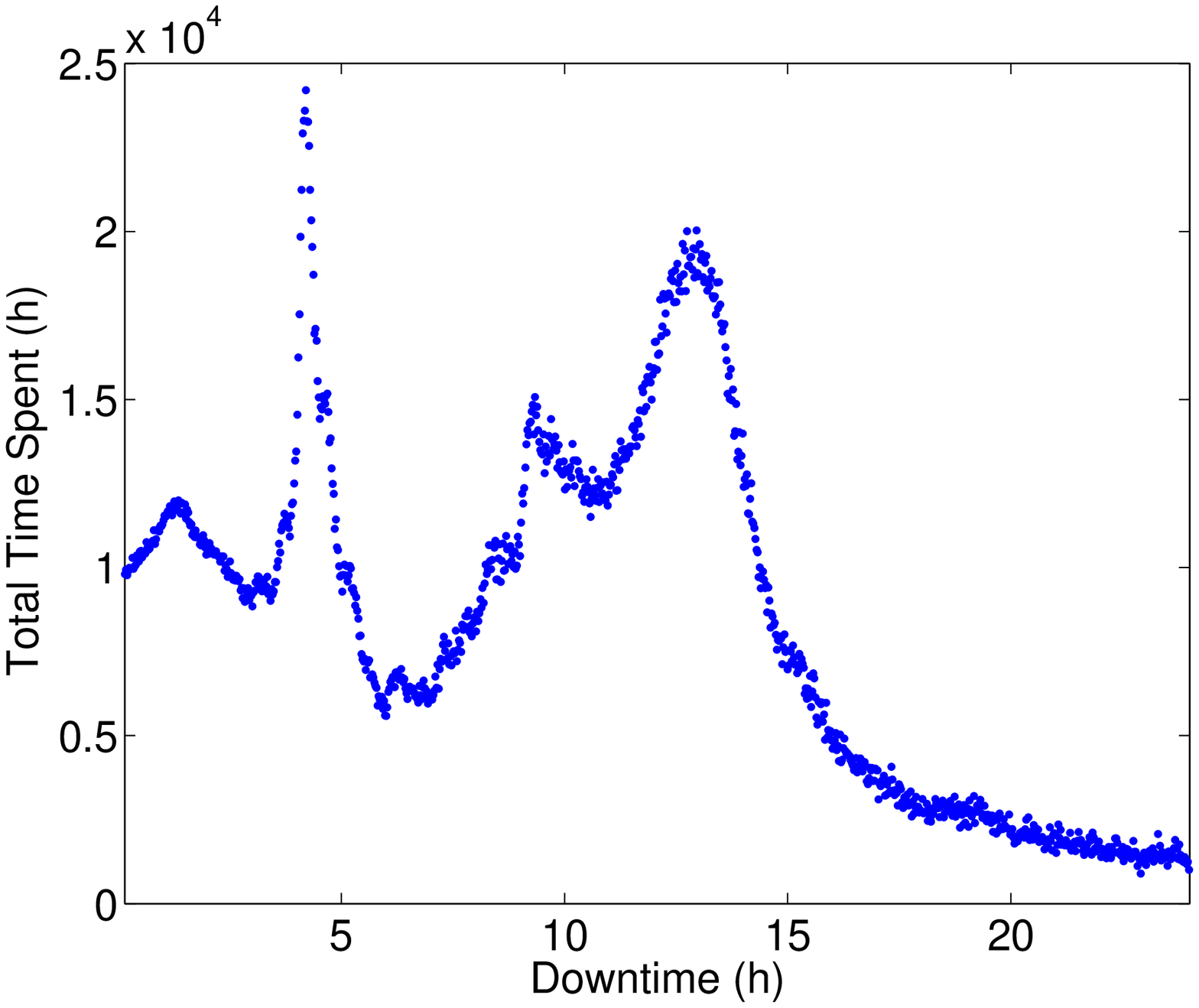,width=6 truecm}} \vspace*{8pt}
\caption{(Left picture) Statistical distribution of the activity
times computed using GPS data of the whole Emilia Romagna region
(blue dots). The straight line suggests the existence of a Benford's
law $p(t)\propto 1/t$. (Right picture) Total activity time
distribution (cfr. eq. (\ref{time2})). The different peaks can be
associated to the main individual activities: part-time job, full
time job and the night rest.\label{time1}}
\end{figure}

The mobility demand is strictly linked to the individual activities.
GPS data do not give information on the individual activities, but
we may assume that each time a driver leaves the engine off more
than 5 minutes, this can be associated to a performed activity (i.e.
the stop is the result of a decision and not accidental). So, we can
study the activity time distribution to understand how individuals
use their time. The result is plotted in the figure \ref{time1}
where we point out the existence of a Benford's
law\footnote{Benford's law is a probability distribution
$P(n)=\ln(1+1/n)$ where $n$ is integer, that can be associated to a
probability density $p(t)=1/t$ where $t$ is real.} $p(t)\propto
1/t$\cite{pietronero2001} that accurately explains the distribution
for $t\le 3$ h ($\simeq 95\%$ of the data): a numerical
interpolation of the experimental data gives $p(t)\propto
1/t^\alpha$ with $\alpha=1.02 \pm 0.02$.
We remark that the empirical Benford's law for the spent time in the
visited locations suggested by the data, is not consistent with the
analogous distributions computed from the mobile phone data
$p(t)\propto t^{-\beta}$ with $\beta=1.8$; this can be the
consequence of the finer time resolution of the GPS data, that
allows to properly consider short time activities. The existence of
a Benford's law for the rest time distribution could be an
indication of a log-time perception(Weber–Fechner
law\cite{dehaene2003}). In other words time is spent proportionally
to the time at disposal\cite{bazzani2010}. The GPS data suggest that
distribution in fig. \ref{time1} is robust and does not depend on
the spatial scale considered (we have the same distribution
considering different cities). To extract relevant information we
consider the distribution $\pi(t)$ of the average time spent for
activities with a time cost $t$ (fig. \ref{time1} right).
\begin{equation}
\pi(t)=tp(t) \label{time2}
\end{equation}
Fig. \ref{time1} shows the peaks related to the main human
activities: the part time job (rest time $t\simeq 4$ h), the full
time job (rest time $t\simeq 8$ h) and the night rest. A small peak
is also around $t\simeq 1.5$ h.\par
\begin{figure}[h]
\centerline{\psfig{file=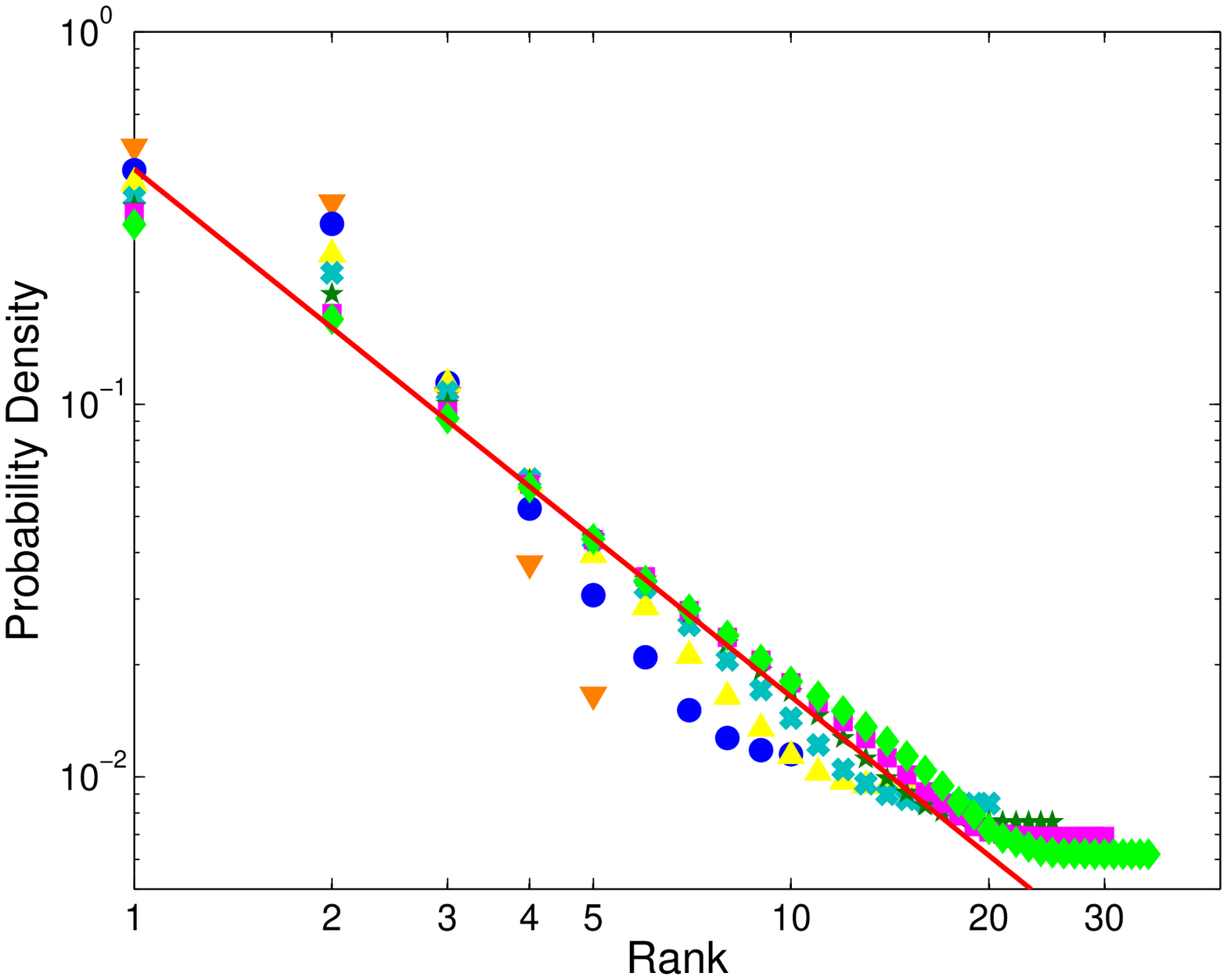,width=6.truecm}
\quad \psfig{file=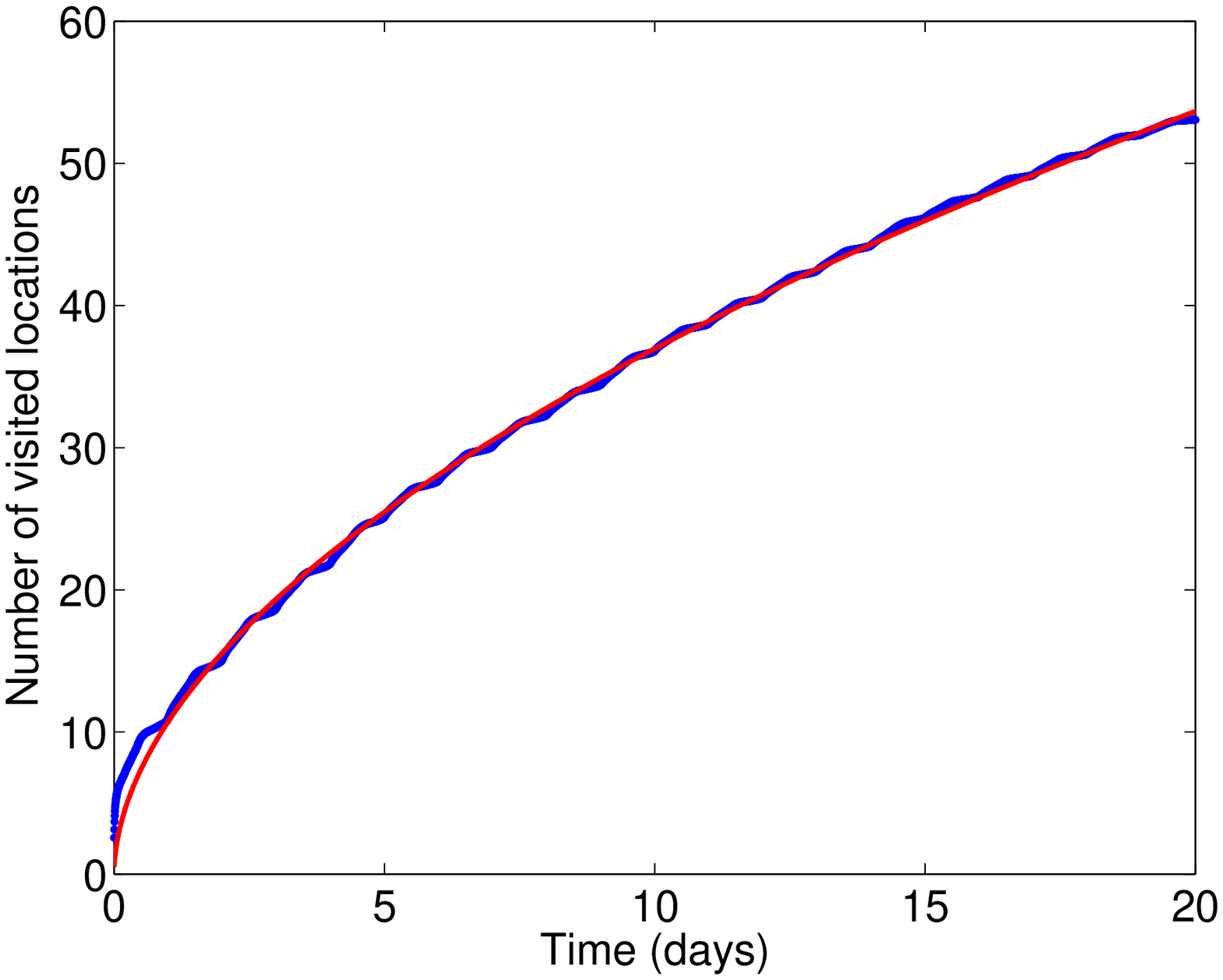,width=6 truecm}}
\vspace*{8pt} \caption{(Left picture) Rank distribution of the
average visitation frequency for each nodes in the individual
mobility networks with a fixed number of nodes: $N=5$ (triangles
down), $N=10$ (circles), $N=15$ (triangles up), $N=20$ (light
squares), $N=25$ (stars), $N=30$ (squares) and $N=35$ (diamonds).
The red line corresponds to a power law interpolation where
$f_k\propto k^{-\alpha}$ where $\alpha=1.42\pm .06$. (Right picture)
Number of visited distinct locations as a function of time (day
unit); the continuous line is an interpolation using a power law
$n(t)\propto t^\beta$ where $\beta=0.5357 \pm 0.006$. In both
figures the data refer to individuals that perform at least 20
mobility days in the considered period. \label{rankf}}
\end{figure}

\noindent To study individual
agenda, we introduce individual mobility networks, where each node
is a visited location and each weighted and directed link implies
the existence of one or more trips between two locations\footnote{we
have clustered the locations whose distance is less than 500 m, that
is an acceptable walking distance from the parking place to the
final destination\cite{benenson2008}.}. Then we have ordered the
nodes from the highest degree to the lowest one for each individual
mobility network (rank distribution). Finally by grouping the
individual mobility networks according to the number of nodes, we
have computed the average visitation frequency $f_k$ of the nodes
with the same rank. The results are reported in the figure
\ref{rankf} where we point out a possible interpolation with a power
law distribution $f_k\propto k^{-\alpha}$ where $\alpha=1.42$.
The existence of a power law distribution for the frequency rank
indicates as the individual activities network is structured
according to a preferential attachment rule, where the most visited
locations could be related to habit mobility. The exponent
$\alpha\simeq 1.42$ computed numerically is in agreement with the
analogous results on human mobility based on a different data
set\cite{song2010}. Under this point of view human mobility can be
represented as a dynamical process on a weighted network where each
individual jumps from one node to another in a random way with
preferential attachment\cite{song2010}. We have empirically studied
the diffusion property of this process using the time dependence of
the total number of different visited locations $n(t)$ for
individuals whose mobility covers at least 20 days in the analyzed
period: i.e. $n(t)$ is the number of new locations visited by the
ensemble of individuals in a time $t$. We apply a Markov hypothesis
to describe the evolution of $n(t)$. Letting $p(k,t)$ be the
probability that the individuals have visited $k$ locations after
$t$ days, we have the Master equation
\begin{equation}
p(k,t+1)=p(k,t)\bar{p}_k+p(k-1,t)(1-\bar{p}_{k-1}) \label{master}
\end{equation}
where $\bar{p}_k$ is an average probability to choose one of the $k$
visited locations, that we assume not dependent from $t$ (stationary
process). By definition
$$
n(t)=\sum_k k p(k,t)
$$
so that from the equation (\ref{master}) we get
\begin{equation}
n(t+1)=n(t)+1-\sum_k p(k,t)\bar{p}_k \label{ave1}
\end{equation}
where we have used the normalizing condition
$$
\sum_k p(k,t)=1
$$
and we have neglected the boundary effect of a finite number of
locations. To proceed, we need to estimate $\bar{p}_k$. Assuming
that individuals perform a Markov's dynamics, $\bar{p}_k$ is the
measure of the $k$ visited locations. The average visitation
frequency $f_k$ can be interpreted both as a measure or as a choice
probability of the $k$ location. Let us order the $k$ locations
according to their rank, the average measure of the $j\in [1,k]$
choice (after $j-1$ choices), $m_j$, can then be estimated
\begin{equation}
m_j\propto \int_j^N f_l^2 dl\propto \int_j^N
\frac{1}{l^{2\alpha}}dl\propto \frac{1}{j^{2\alpha-1}}\qquad N\gg 1
\label{est1}
\end{equation}
where we have used the power law interpolation for the rank
distribution $f_k\propto k^{-\alpha}$. As a consequence, the
expected measure for the $k$ visited locations reads
\begin{equation}
\bar{p}_k \propto\sum_{j=1}^k \frac{1}{j^{2\alpha-1}}\simeq \int_1^k
\frac{1}{j^{2\alpha-1}}dj\propto \left (1-
\frac{1}{k^{2\alpha-2}}\right ) \label{est2}
\end{equation}
By definition $\bar{p}_k\to 1$ as $k$ increases. Using the estimate
(\ref{est2}), the equation (\ref{ave1}) reads
\begin{equation}
n(t+1)=n(t)+1-\sum_k p(k,t)\left (1- \frac{1}{k^{2\alpha-2}}\right
)=n(t)- \sum_k \frac{p(k,t)}{k^{2\alpha-2}}\label{ave2}
\end{equation}
Then we apply the mean field theory argument to reduce the equation
(\ref{ave2}) to the simple form
\begin{equation}
n(t+1)=n(t)+\frac{1}{n(t)^{2\alpha-2}} \label{aveq}
\end{equation}
whose solution can be approximated by
\begin{equation}
n(t)\simeq c t^{1/(2\alpha-1)} \label{nt}
\end{equation}
where $c$ is an integration constant. According to the $f_k$
interpolation (fig. \ref{rankf} left) $\alpha\simeq 1.42\pm .06$ and
we get
$$
n(t)\propto t^\beta
$$
where $\beta=.54\pm .03$. The result is very close to the numerical
interpolation of the empirical measures, that gives $\beta=.53$
(fig. \ref{rankf} right) and it has to be compared with the result
$\beta=.60\pm .02$ reported in the literature using mobile phone
data\cite{song2010}. But we have to remark that the model presented
in the paper\cite{song2010} to reproduce the individual mobility
cannot be applied in our case since it is not consistent with the
empirical activity time distribution (fig. \ref{time1}).  The
results may be interpreted in a twofold way. From one hand this is
another indication that macroscopic statistical properties of human
mobility mimics the properties of an ensemble of particles which
perform a stochastic Markov dynamics, taking into account the
existence of spatial and temporal constraints. On the other hand the
individual dynamics is certainly not a Markov process and the rank
distribution in fig. \ref{rankf} is the result of a cognitive
behavior that defines the daily mobility agenda in a complex urban
environment.

\section{Conclusions}

We have analyzed some statistical properties of human mobility, that
are related to the use of private cars. Our results are based on a
sample of individual trajectories ($\simeq 2\%$) of the whole
vehicle population of an Italian region (Emilia Romagna) (22000
km$^2$) recorded using a GPS system. We have shown as the daily
mobility time can be used as ''cost function'' to describe the
average behavior of individuals in a stationary situation, in a
theoretical framework similar to the Boltzmann's gas model. An
exponential-like distribution is also suggested by the individual
classification according to the average number of daily trips. The
empirical distribution of the single trip length up to $15$ km and
more, is consistent with a Boltzmann'gas framework if one assumes a
random model for the choice of the trip length given the daily
mobility and the number of activities. Our analysis points out that
the short trip distribution, that is mainly related to urban
mobility, does not obey to a power law, but closely obeys to
exponential law. The empirical data suggests only the existence of a
power law behavior for the long trip distribution, that could be
interpreted by means of a correlation between the number of
performed activities and the daily mobility. To recover an empirical
power law distribution for individual mobility, as reported in the
literature\cite{brockmann2006}, one has probably to consider
different transportation means. The study of the downtime
distribution between two successive trips has confirmed the
existence of a universal Benford's law\cite{pietronero2001} that
could be related to a log-time perception\cite{dehaene2003}, when
individuals perform their daily ''asystematic'' activities. Whereas
the three peaks detected in the average time distribution correspond
to the systematic activities of part-time and full-time jobs, and
the night rest. Our analysis points out that the majorities of the
performed activities seems to obey to a statistical distribution of
independent events even if the time spent in the ''systematic
activities'' is relevant. This may indicate that human behavior
during mobility has strong stochastic components and the
predictability of the spatial displacements can be difficult,
contrarily to the localization in time\cite{song2010b}. Finally, we
have performed a study to unroll the relevance of habits in the
individual mobility. Our results are in a qualitative agreement with
analogous results based on different data and confirm the idea that
the individual mobility networks can be understood using a
preferential attachment paradigm, but the comprehension of the
primal mechanisms of the individual mobility demand is still an open
problem.
\section*{Acknowledgments}
We thanks Octo Telematics s.p.a. for providing the GPS database.

\end{document}